\documentclass[fleqn,twoside]{article}
\usepackage[dvips]{graphicx}
\usepackage[dvips]{pstcol} 
\usepackage{espcrc2}
\newcommand{\be}{\begin{equation}}
\newcommand{\en}{\end{equation}}
\newcommand{\bea}{\begin{eqnarray}}
\newcommand{\ena}{\end{eqnarray}}
\newcommand{\lapprox}{%
\mathrel{%
\setbox0=\hbox{$<$}\raise0.6ex\copy0\kern-\wd0\lower0.65ex\hbox{$\sim$}}}

\newcommand{\gapprox}{%
\mathrel{%
\setbox0=\hbox{$>$}\raise0.6ex\copy0\kern-\wd0\lower0.65ex\hbox{$\sim$}}}
\newcommand{\mpid}{m^2_\pi}

\newcommand{\mpd}{{m^2_+}}

\def\im{ {\rm Im}\,}
\def\upi{ {1\over\pi}}
\def\udemi{ {1\over2}}
\def\tdemi{ {3\over2}}
\def\utier{ {1\over3}}

\def\qtier{ {4\over3}}

\setbox4=\hbox{$-$}
\def\pvint#1#2{\copy4\kern-1.1\wd4\hbox{$\displaystyle\int_{#1}^{#2}$}}
\def\xlss{\lambda_{s'}}

\def\xls{\lambda_{s}}

\definecolor{LightGray}{rgb}{0.92,0.92,0.92}
\definecolor{Gray}{rgb}{0.75,0.75,0.75}
\definecolor{VeryLightBlue}{rgb}{0.9,0.9,1}
\definecolor{LightBlue}{rgb}{0.6,0.6,1}
\definecolor{DarkBlue}{rgb}{0,0,0.6}
\definecolor{LightGreen}{rgb}{0.88,1,0.88}
\definecolor{MidGreen}{rgb}{0.6,1,0.6}
\definecolor{DarkGreen}{rgb}{0,0.6,0}
\definecolor{VeryLightYellow}{rgb}{1,1,0.9}
\definecolor{LightYellow}{rgb}{1,1,0.6}
\definecolor{MidYellow}{rgb}{1,1,0.5}
\definecolor{DarkYellow}{rgb}{1,0.8,0.0}
\definecolor{VeryLightRed}{rgb}{1,0.9,0.9}
\definecolor{LightRed}{rgb}{1,0.8,0.8}
\definecolor{Brown}{rgb}{0.9,0.5,0.0}
\definecolor{Lemon}{rgb}{1.0,0.98,0.8}
\definecolor{Red}{rgb}{1.,0.,1.}
\newcommand{\rouge}[1]{\textcolor{red}{#1}}

\newcommand{\verte}[1]{\textcolor{DarkGreen}{#1}}

\title{$\pi K$ scattering inputs to ChPT\thanks{
Work supported in part by the EU RTN contract HPRN-CT-2002-00311 (EURIDICE).
Talk presented by B.M. }}

\author{
Paul B\"uttiker\address{ Theory division of the Helmholtz Institute,
University of Bonn, Germany}
S\'ebastien Descotes-Genon\address{Laboratoire de Physique Th\'eorique,
Universit\'e Paris-Sud, 91406 Orsay, France}
Bachir Moussallam\address{ Institut de Physique Nucl\'eaire, 
Universit\'e Paris-Sud, 91406 Orsay, France} }

\begin{document}
\begin{abstract}
Experimental information on low energy $\pi K$ scattering would shed
light on the poorly known OZI suppressed sector of ChPT. I describe 
recent work aimed at 
generating such information based on available experimental data
by setting up and then solving with appropriate boundary conditions
a non linear system of equations of the Roy and Steiner type. First
results of this analysis are presented.
\end{abstract}
\maketitle

This talk describes work done with Paul B\"uttiker and S\'ebastien 
Descotes-Genon.
Following ideas and methods due to Roy\cite{roy} and to Steiner\cite{steiner}, 
our aim was to generate data on the  $\pi K$ scattering amplitude
at very low energy, even in unphysical regions, using input experimental data
at medium and high energy.
I will begin by explaining the need for such $\pi K$ data in connection
with developments of ChPT and questions about different chiral limits.

The fundamental degrees of freedom of QCD are quarks 
and gluons. Because the three lightest quarks happen
to have masses much smaller than  1 GeV
it is possible, in a limited kinematical region of 
the non perturbative regime, 
to make a change 
of variables and use an effective Lagrangian where $\pi$, $K$ and $\eta$
are the fundamental degrees of freedom. 
This Lagrangian allows one to
perform expansions around chiral limits. There are two different 
chiral limits which are relevant to the physical world:
one with $N^0_F=2$ massless flavours and one with $N^0_F=3$.
The chiral Lagrangian at order $p^4$ (NLO) was constructed by Gasser and
Leutwyler\cite{gl85} and involves 10 independent coupling constants $L_i$.
More recently, the chiral Lagrangian at NNLO was constructed\cite{bceg}
which brings in a number of new couplings $C_i$. It becomes of obvious
importance to collect as much experimental data as possible to test the
theory.
The pion-Kaon scattering
amplitude turns out to be a particularly interesting process in connection
with ChPT. The computation of the amplitude at order $p^4$ was performed 
in ref.\cite{bkm}. By matching this expression with low energy experimental 
input allows one to probe many of the
chiral couplings including those which are suppressed in the  large 
$N_c$ limit, like the coupling $L_4$ \cite{abm}.
Such couplings ($L_4$, $L_6$ and the combination $L_2-2L_1$ )
were set equal to zero (and a plausible guess of the 
error was made) in the original work or ref.\cite{gl85}.
Their actual values have interesting physical implications.
For instance, the value of $F_\pi$ in a chiral limit is an order parameter
for the spontaneous breaking of  chiral symmetry. 
The coupling $L_4$ controls  the difference 
in the values of $F_\pi$ in the chiral limits with $N^0_F=2$
and $N^0_F=3$. This difference is a non trivial dynamical property of QCD 
which is linked to the puzzling properties of the light scalar mesons.
An unrealistic but amusing illustration of this link is provided by the
``extended'' linear sigma model discussed in ref.\cite{mou00}: for certain
scalar meson assignments a dramatically large change in $F_\pi$ is predicted. 
The low energy $\pi K$ amplitude probes not only $L_4$ but also to some extent
$L_6$ (via the combination $L_8+2L_6$ ) and also several other chiral coupling
constants.
Experimental data on the  $\pi K$  amplitude
at sufficiently low energies are either unavailable or unreliable 
but one can construct such data based on reliable experimental input.
Such a construction is made possible because the S-matrix
has analyticity properties, from which one can write down dispersion 
relations. Next, the property of crossing allows one to determine the 
subtraction functions. What makes the pseudo-scalar mesons unique in the
application of these methods is that they are the lightest particles
in the QCD spectrum. 
As a consequence, scattering of pseudo-scalar mesons is elastic at low
energies. In practice, 
there exists a significant
energy region ($E\lapprox $ 1 GeV ) in which scattering can be considered as
elastic to a very good approximation. 
Finally, in this
same region the partial waves with $l\ge 2$ are negligibly  small such that
after
projecting the dispersion relations over partial waves one obtains a closed
system of equations for the S and the P waves in the energy region
$E\le E_m$ (with $E_m\simeq $ 1 GeV is called the {\sl matching point}).  
The amplitude for $E \ge E_m$  must be provided as input to
these equations. 
Such equations were first proposed
by Roy\cite{roy} (for $\pi\pi$ scattering) and by Steiner\cite{steiner}. 
While $\pi\pi$ scattering
was intensively studied (see e.g.\cite{acgl} and references therein), 
much less work was devoted to $\pi K$.
Also, in earlier work\cite{bonnier,lang,johannesson} no accurate
experimental input data was available. 
In the $\pi K$ case, one can derive a set of six coupled equations 
which involve the four
partial waves $f_0^I$,$f_1^I$ with $I={1\over2},\ {3\over2}$ 
of the $\pi K\to\pi K$ amplitude and the two partial waves
$g_0^0$, $g_1^1$ of the $\pi\pi\to K\overline{K}$ amplitude. 
These equations contain two 
arbitrary parameters which are conveniently chosen to be 
the two S-wave scattering lengths, $a_0^{1/2}$, $a_0^{3/2}$. A typical
equation is shown below
\bea
&&\verte{Re f_l^\udemi(s)}
= k^\udemi_l(\rouge{a_0^{1/2},a_0^{3/2}},s) 
+\upi\pvint\mpd{\infty} ds'\,
\sum_{l'=0,1}
\nonumber\\&&
\Bigg\{ \left(\delta_{ll'}{\xls\over(s'-s)\xlss}
-\utier K^\alpha_{ll'}(s,s')\right)
\verte{\im f_{l'}^\udemi(s')}
%\nonumber\\ &&\phantom{Re f_l^\udemi(s)= k^\udemi_l(s) 
%+\upi\pvint\mpd{\infty} ds'\,\sum_{l'=0,1}}
\nonumber\\&&
+\qtier K^\alpha_{ll'}(s,s') 
\verte{\im f_{l'}^\tdemi(s')} \Bigg\}
+\upi\int_{4\mpid}^{\infty} dt' 
\nonumber\\&&
\Big\{ K^0_{l0}(s,t') 
\textcolor{magenta}{\im g_0^0(t')}
+2 K^1_{l1}(s,t') 
\textcolor{magenta}{\im g_1^1(t')}\Big\}
\nonumber\\&&
+d^\udemi_l(s)
\ena
which displays the usual singular Cauchy kernel and other non-singular
kernels. Elastic unitarity provides an additional non-linear relation
between real an imaginary parts.

Important progress was achieved recently by Gasser and 
Wanders\cite{gasser-wanders}
in clarifying the multiplicity properties of the Roy equations solutions. 
The $\pi K$ equations can be recast in a form analogous to that considered
in ref.\cite{gasser-wanders} after eliminating $g_0^0$, $g_1^1$. 
The multiplicity index is controlled by
the values of the $\pi K$ phase shifts at the matching point.
At the matching point, appropriate boundary conditions must be enforced:
firstly, one must impose that the phase shifts are continuous.
In practice this condition cannot be applied to $\delta_1^{3/2}$ which is
too small and inaccurately measured at the matching point. This P-wave must
be treated on the same footing as the $l\ge2$ partial waves and does not
influence the multiplicity index. 
Choosing the matching point to
be at 1 GeV approximately, the multiplicity index turns out to vanish.
This implies that for a given set of values for $a_0^{1/2}, a_0^{3/2}$
if a solution exists, then it is unique.
Two additional physical 
requirements that one can impose are that the derivatives
of two of the phase shifts should also  be continuous. 
These additional constraints can no longer be satisfied for arbitrary values of
$a_0^{1/2}, a_0^{3/2}$. If the  input data were perfect, the S wave scattering
lengths would be exactly  determined 
as discrete eigenvalues of the system of Roy-Steiner (RS) equations together
with the appropriate boundary conditions.  
Our work has consisted in determining how the S wave scattering lengths are
constrained in practice from the available experimental data and their errors.
\begin{figure}[h]
\includegraphics[width=8cm]{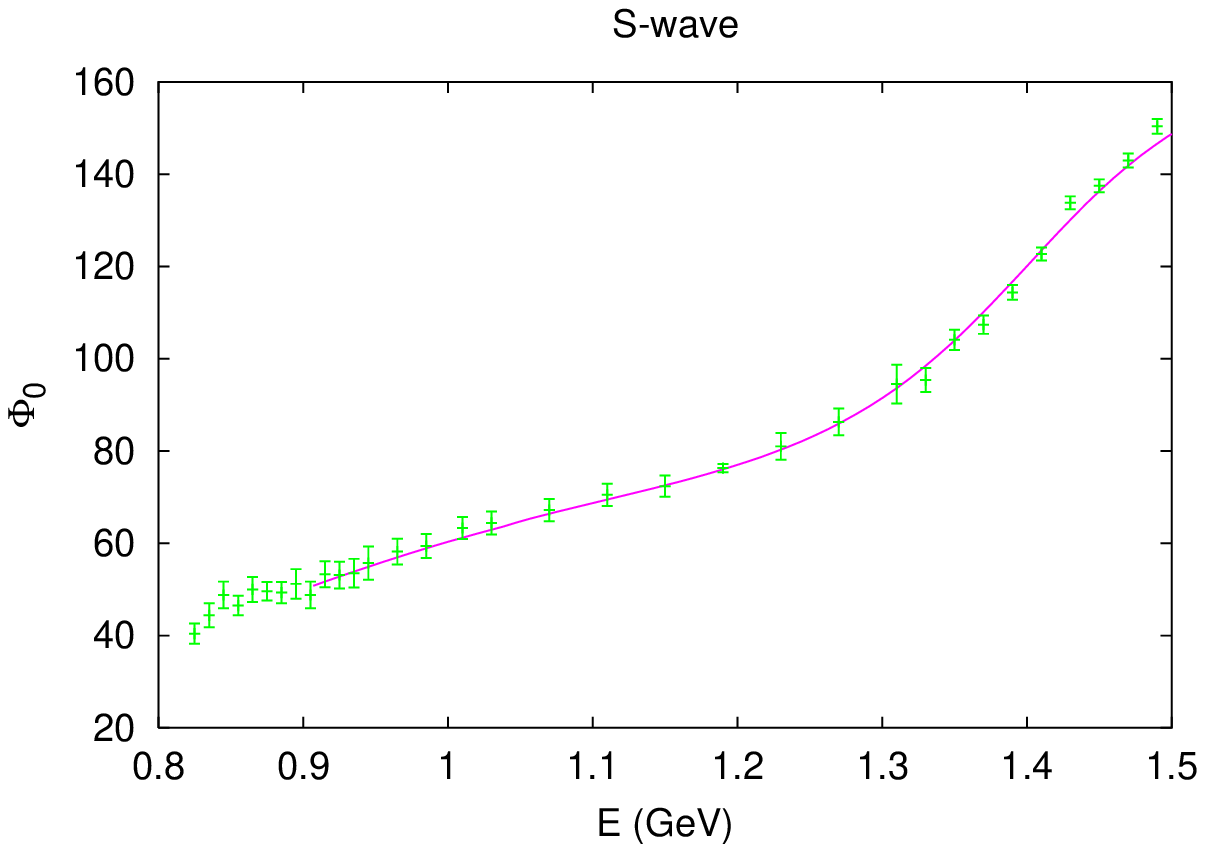}
\includegraphics[width=8cm]{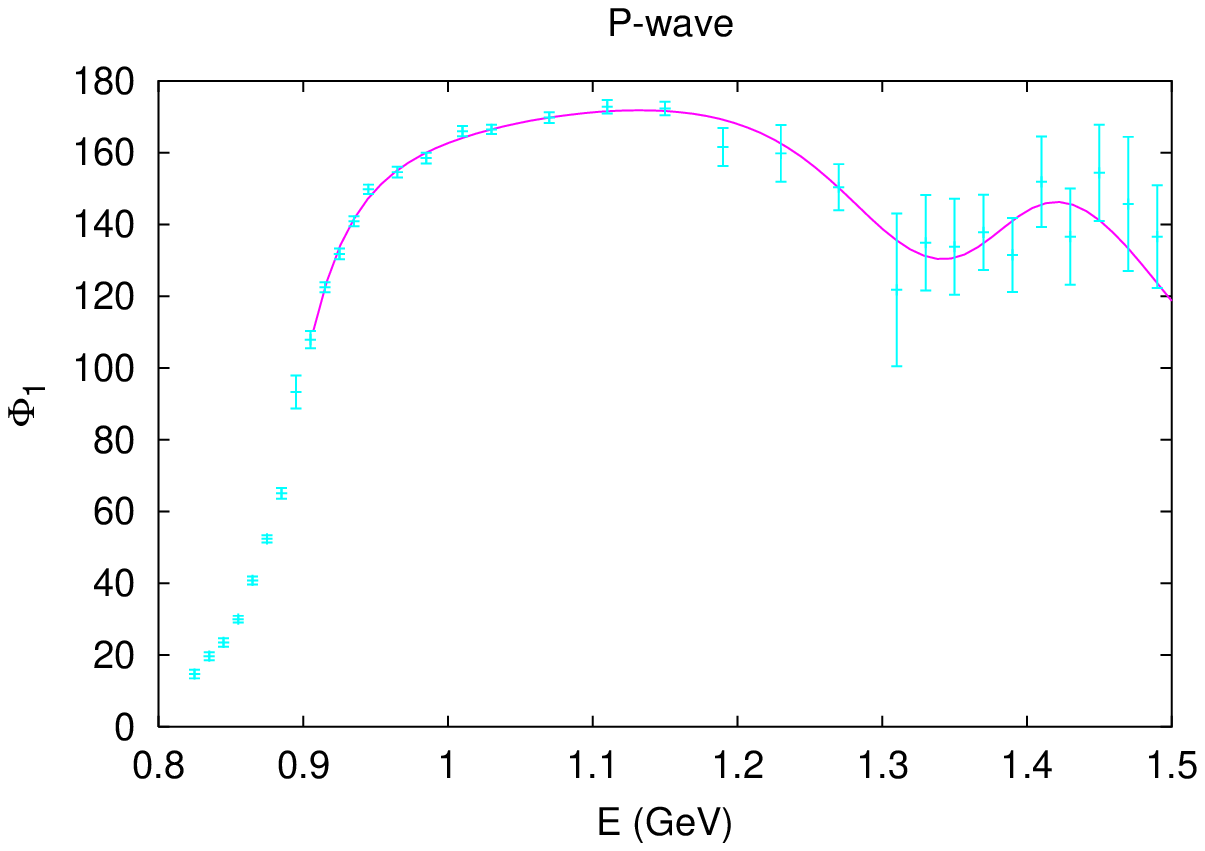}
\caption{\sl Input data for the $I=1/2$ S- and P-waves. }
\end{figure}

Let us now examine the experimental input data. In our analysis, we have used
the experimental data for $\pi K\to\pi K$ 
from Estabrooks et al.\cite{estabrooks} and from Aston et al.\cite{aston} 
which are
both high statistics production 
experiments considerably more accurate than the data
which were available before. We also need data on the $\pi\pi\to K\overline{K}$
amplitudes. For these, we have used the results from Cohen et al.\cite{cohen} 
and from Etkin et al.\cite{etkin} which are also generated from 
high-statistics production experiments. We note that the amplitudes in the
unphysical region below the $K\overline{K}$ threshold are generated from
solving the RS equations.
Fig. 1 displays the data for the $\pi K$ S- and the P-waves
it shows that the data are rather smooth and accurate in the region 
of the matching points. 
In order to ascertain
the values of the phases and of the derivatives at the matching point
(which are crucial in that analysis) we have performed several different
types of fits. We use much more input than showed here: like higher partial
waves, Regge models for asymptotic regions etc... 
more details will be provided in a forthcoming publication.
The RS equations are then solved to high
numerical accuracy. 

The errors on the output of the equations are estimated
by varying the parameters used in the fits to the input data and making use
of their correlation matrices.
\begin{figure}[h]
\includegraphics[width=8cm]{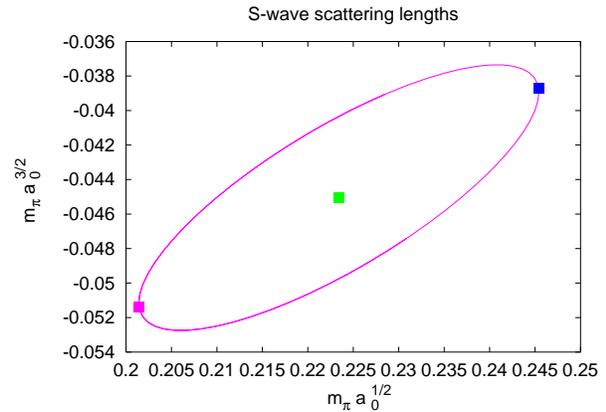}
\caption{\sl One-sigma ellipse for the two S-wave scattering lengths
obtained by solving the RS equations with appropriate boundary conditions.}
\end{figure}
The main result of this analysis is the determination of a region 
inside which the two S-wave scattering lengths are constrained. 
This one-sigma ellipse is shown in Fig. 2. 
This result is rather non trivial as no experimental data at all 
has been used below one GeV ! 
As one can see from the figure, this region is rather small: this is a direct
reflection of the precision of the data used as input, in particular in 
the region of the matching point. 
\begin{figure}[h]
\includegraphics[width=8cm]{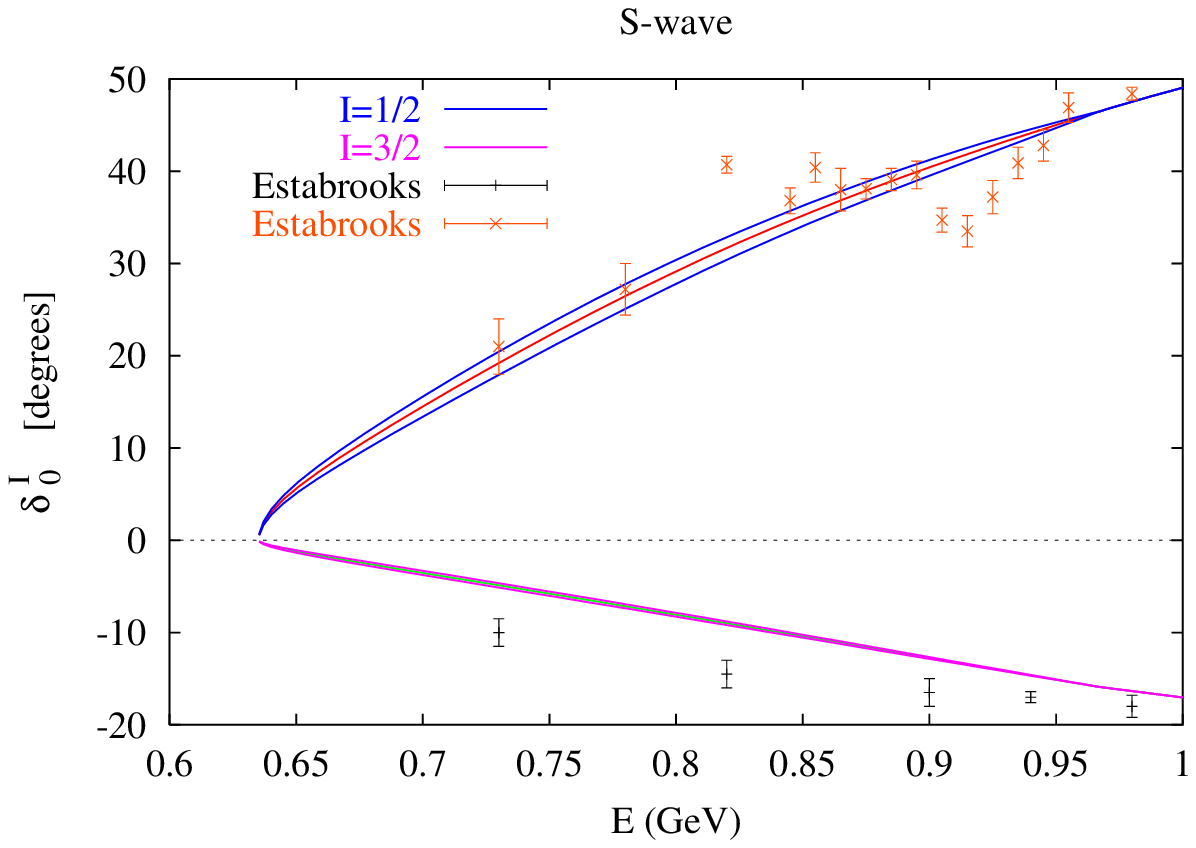}
\includegraphics[width=8cm]{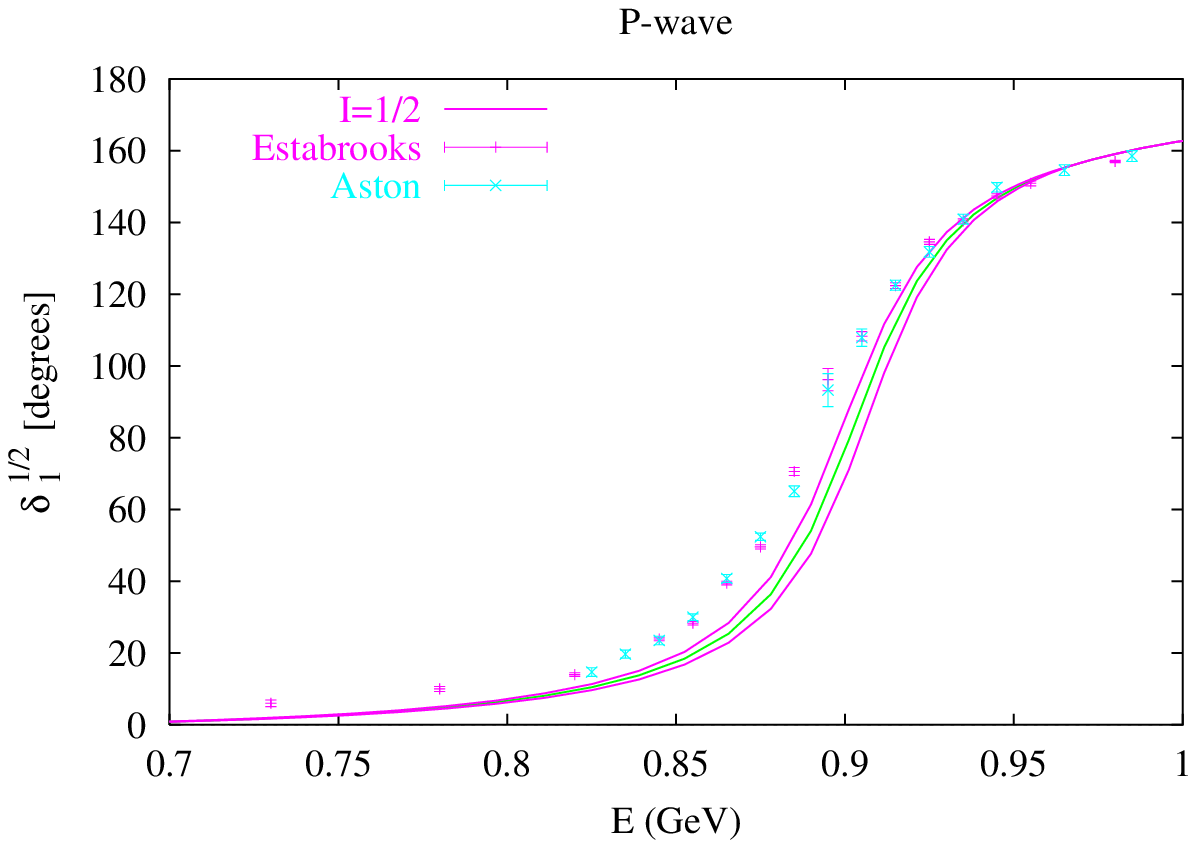}
\caption{\sl Results for S and P-wave phase-shifts in the low energy region
obtained from solving the RS equations.}
\end{figure}
Fig. 3 shows  phase shifts that we obtain from solving
the RS equations. 
We display in every case three solutions corresponding to the
three points in the $a_0^{1/2}$ $a_0^{3/2}$ plane as shown in Fig. 2. 
We also show for comparison some of the experimental data available
in this region. A priori, the determination of the phase shifts from production
experiments become less reliable at lower energies. In general, our results
are not in very good agreement with the data in this region. The most striking
disagreement concerns the P-wave. Our prediction for the mass of the $K^*$
is higher by 10 MeV than the mass from refs.\cite{estabrooks,aston}. One 
possible reason for this discrepancy is isospin breaking which is not 
properly taken into account in our analysis but we also note that CLEO
has reported a discrepancy between the mass of the $K^{*+}$
found in $\tau\to \pi K_S \nu$
decays and the PDG value\cite{urheim}. 

Using the results of our work we can calculate
the $\pi K$ amplitude at the threshold or below the threshold and 
match with ChPT expansions. These results will be presented 
elsewhere. In the future, we expect
new data to become available which would help sharpen our predictions.
Better data on the P-wave phase shift from  CLEO is one example. 
Reliable data on the S-wave phase shifts could be provided from the $D$
decay mode $D\to \pi K \mu \nu$. Finally, $\pi K$ atom experiments are 
planned which could directly access combinations of scattering lengths.

\end{document}